\documentclass[prb,aps,tightenlines,twocolumn,
showpacs,floatfix]{revtex4}
\usepackage{graphicx}
\usepackage{amssymb}
\usepackage{amsmath}
\usepackage{bm}
\usepackage{colordvi}
\def\up{\uparrow}
\def\down{\downarrow}

\begin{document}
\title{Spin relaxation due to random Rashba spin-orbit coupling in
  GaAs (110) quantum wells} 
\author{Y. Zhou}
\author{M. W. Wu}
\thanks{Author to whom correspondence should be addressed}
\email{mwwu@ustc.edu.cn.}
\affiliation{Hefei National Laboratory for Physical Sciences at
  Microscale and Department of Physics, University of Science and
  Technology of China, Hefei, Anhui, 230026, China}

\date{\today}
\begin{abstract}
We investigate the spin relaxation due to the random Rashba spin-orbit
coupling in symmetric GaAs (110) quantum wells from the fully
microscopic kinetic spin Bloch equation 
approach. All relevant scatterings, such as the electron-impurity,
electron--longitudinal-optical-phonon, electron--acoustic-phonon, as
well as electron-electron Coulomb scatterings are explicitly included.
It is shown that our calculation reproduces the experimental data by 
M\"uller {\em et al.} [Phys. Rev. Lett. {\bf 101}, 206601 (2008)]
for a reasonable choice of parameter values.
We also predict that the temperature dependence of spin relaxation
time presents a peak in the case with low impurity density, which
originates from the electron-electron Coulomb scattering. 
\end{abstract}


\maketitle

Semiconductor spintronics has been an active field of research lately due to
the potential application of spin-based devices.\cite{opt-or,Wolf,spintronics}  
Recent experiments show that the spin relaxation time (SRT) in
(110)-oriented GaAs quantum wells (QWs) is extremely long, and thus
the spin dynamics in this system has attracted much attention both
experimentally and theoretically.\cite{Ohno_99,Ohno_2001,Wu_110,
Dohrmann_04,Hagele_05,gate1,gate2,AC_110_1,AC_110_2,sysmetry1,
sysmetry2,Muller_08,YZhou_SSC_09,Tarasenko,Cartoixa} 
The physics underlying this effect is the absence of the
D'yakonov-Perel' (DP) mechanism,\cite{Dyakonov} which is 
the leading spin relaxation mechanism in $n$-type zinc-blende
semiconductors. 
The DP mechanism is from the joint effects of the momentum scattering
and the momentum-dependent effective magnetic field (inhomogenous
broadening\cite{wu_review}) induced by the
Dresselhaus\cite{Dresselhaus_55} and the Rashba\cite{Rashba_84} 
spin-orbit coupling (SOC). In symmetric GaAs (110) QWs with only the
lowest subband occupied, the in-plane component of the
spin-orbit field vanishes.\cite{Winkler_04} 
Therefore the DP mechanism cannot affect electrons with spin
polarization along the growth direction and the SRT in this
system is considerably larger than that in (100) QWs. 
In most of the previous works, the main reason limiting the SRT 
is attributed to the Bir-Aronov-Pikus 
mechanism,\cite{Bir} which is from the exchange interaction
between the electrons and the photo-generated holes.
One of the exceptions is the spin noise spectroscopy measurement by
M\"uller {\em et al.},\cite{Muller_08} where the excitation of
semiconductor is negligible and hence the Bir-Aronov-Pikus mechanism is
avoided. They reported the longest SRT in this system which is about 
$24$~ns. Since the DP and Bir-Aronov-Pikus mechanisms are both absent, and the
virtual intersubband spin-flip SOC induced spin relaxation is also
ruled out due to the relatively high mobility of the
samples,\cite{YZhou_SSC_09} 
the possible reason limiting the SRT is the DP mechanism
due to the random Rashba SOC caused by the fluctuations of the donor
density.\cite{Sherman_03,Sherman_PRB_05} 

As shown by Sherman {\em et al.},\cite{Sherman_03,Sherman_PRB_05}  
even in symmetric QWs, the unavoidable fluctuations of the
concentration of the dopant ions still lead to a random electric 
field along the growth direction, and hence a random Rashba
SOC at each point of a QW. This random SOC provides
an inhomogeneous broadening and  induces the DP spin
relaxation. 
The previous investigations\cite{Sherman_03,Sherman_PRB_05} on the
spin relaxation due to this mechanism are based on the single-particle
theory, and thus the electron-electron Coulomb scattering, which has
been shown to be very important for spin relaxation in two-dimensional
system,\cite{wu_review,Weng_04,Zhou_PRB_07,wu-exp,Ivchenko,Leyland,Ruan,Teng}
is missing. In this work, we apply the fully microscopic kinetic spin Bloch
equation (KSBE) approach\cite{wu_review,Zhou_PRB_07} to investigate the
spin relaxation due to the random Rashba SOC in symmetric GaAs (110)
QWs. Here all relevant scatterings, especially the electron-electron
Coulomb scattering, are explicitly included. 
We will show that our calculation is in good agreement with the
experimental data by M\"uller {\em et al.}.\cite{Muller_08}
We also predict that the temperature dependence of the SRT presents a peak
due to the electron-electron Coulomb scattering in the
low-impurity-density case.

We start our investigation from an $n$-type symmetrically
modulation-doped GaAs (110) QW with the growth direction along the $z$
axis. The in-plane coordinate axes are set as $x\|[1\bar{1}0]$ and
  $y\|[00\bar{1}]$. A magnetic field $B$ is applied along the
$x$-axis. The well width is small enough so that only the lowest
subband is occupied for the temperature and electron
density we discuss. The envelope function of the relevant subband
is calculated under the finite-well-depth assumption.\cite{Zhou_PRB_07}
The barrier layer is chosen to be Al$_{0.39}$Ga$_{0.61}$As as 
the experiment\cite{Muller_08} where the
barrier height is 319~meV.\cite{Yu_92}
By using the nonequilibrium Green's function method,\cite{Haug_1998}
the KSBEs can be constructed as:\cite{wu_review}
\begin{equation}
  \partial_t\hat{\rho}_{{\bf k}}= \partial_t\hat{\rho}_{{\bf k}}|_
  {\rm coh}+\partial_t\hat{\rho}_{{\bf k}}|_{\rm scat},
\end{equation}
where $\hat{\rho}_{\bf k}$ represents the electron single-particle
density matrix, whose diagonal and off-diagonal elements describe the
electron distribution function $f_{{\bf k}\sigma}$ and spin coherence
$\rho_{\bf k}$ respectively. 
The scattering term $\partial_t\hat{\rho}_{{\bf k}}|_{\rm coh}$
consists of the electron-impurity, electron--longitudinal-optical
(LO)-phonon, electron--acoustic (AC)-phonon and electron-electron
Coulomb scatterings, whose expressions are given in detail in
Ref.~\onlinecite{Zhou_PRB_07}. The coherent term can be written as
($\hbar\equiv 1$ throughout this paper) 
\begin{equation}
\left.\partial_t\hat{\rho}_{{\bf k}}\right|_{\rm coh}=
-i\Big[{\bf h}({\bf k})\cdot \hat{\mbox{\boldmath$\sigma$}}
+\hat{\Sigma}_\mathrm{HF}({\bf k}),\;\; \hat{\rho}_{{\bf k}} \Big],
\end{equation}
in which $[A,B]\equiv AB-BA$ is the commutator. 
$\hat{\Sigma}_\mathrm{HF}({\bf k})$ is the effective magnetic field
from the Coulomb Hartree-Fock contribution.\cite{wu_review}
${\bf h}({\bf k})$ represents the spin-orbit field composed of the
Dresselhaus\cite{Dresselhaus_55} and Rashba\cite{Rashba_84}
terms. The Dresselhaus spin-orbit field in (110) QWs reads 
\begin{equation}
  {\bf h}_{\rm D}({\bf k})=\gamma_D\Big(0,\;0,\;
  \frac{k_x}{2}(k_x^2-2k_y^2- \langle k_z^2\rangle)\Big).
\end{equation}
Here $\gamma_{\rm D}$ denotes the Dresselhaus SOC coefficient and
$\langle k_z^2 \rangle$ stands for the average of the operator
$-(\partial / \partial z)^2$ over the electronic state of the lowest
subband. 
The effective magnetic field from the Rashba SOC can be written as
\begin{equation}
  {\bf h}_{\rm R}({\bf k})=\alpha_{\rm R}(k_y,\;-k_x,\;0)
\end{equation}
with $\alpha_{\rm R}$ representing the Rashba SOC coefficient.
To incorporate the effect of the random Rashba SOC,
we assume that $\alpha_{\rm R}$ satisfies the Gaussian
distribution function $P(\alpha_{\rm R})=\frac{1}{\sqrt{2\pi} 
\Delta_{\rm R}} e^{-\alpha_{\rm R}^2/2\Delta_{\rm R}^2}$ 
with $\Delta_{\rm R}$ being the standard deviation of the distribution 
function.\cite{Lorentz} 
We divide the regime of $\alpha_{\rm R}$ from $-\alpha_{\rm cut}$
to $\alpha_{\rm cut}$ equidistantly by $N_{\rm R}$ odd nodes, where
$\alpha_{\rm cut}$ is the cutoff value of $\alpha_{\rm R}$ regime.
For each node of $\alpha_{\rm R}$, we obtain the temporal
evolutions of the electron distribution function by numerically
solving the KBSEs. The SRT is obtained by the slope of the envelope of
the coherently summed spin polarization 
\begin{equation*}
  S_z=\sum_{\alpha_{\rm R}}\Delta\alpha_{\rm R}P(\alpha_{\rm R})\sum_{{\bf k}}\frac{1}{2}
  [f_{{\bf k}\up}(\alpha_{\rm R}) - f_{{\bf k}\down}(\alpha_{\rm R})]
\end{equation*}
with $\Delta\alpha_{\rm R}=2\alpha_{\rm cut}/(N_{\rm R}-1)$. 
It is checked that $\alpha_{\rm cut}=1.6$~meV$\cdot${\AA} and 
$N_{\rm  R}=15$ is sufficient for the convergence of our calculation.


\begin{figure}[tbp]
  \begin{center}
    \includegraphics[width=6.5cm]{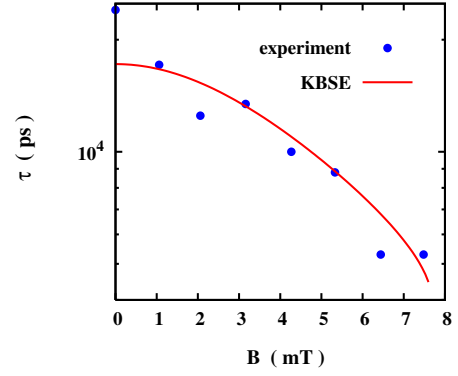}
  \end{center}
  \caption{(Color online) SRTs from the experimental data in
    Ref.~\onlinecite{Muller_08} ($\bullet$) and from the KSBEs (solid
    curve) {\em vs.} the applied magnetic field. Here $T=20$~K, $a=16.8$~nm,
    $N_e=1.8\times10^{11}$~cm$^{-2}$ and $N_i=0.01~N_e$. 
  }
  \label{fig_mag}
\end{figure}

\begin{figure}[htbp]
  \begin{center}
    \includegraphics[width=6.5cm]{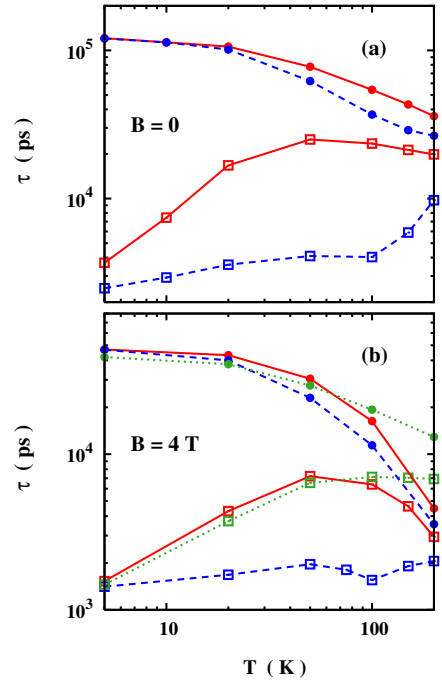}
  \end{center}
  \caption{(Color online) SRTs due to the DP mechanism induced by the
    random Rashba field {\em vs.} temperature $T$ for $B=0$ (a) and
    $4$~T (b) with impurity densities $N_i=N_e$ ($\bullet$) and
    $0.01~N_e$ ($\square$). 
    Here $a=16.8$~nm and $N_e=1.8\times10^{11}$~cm$^{-2}$. 
    The corresponding Fermi temperature is $75$~K. The red solid
    curves represent the results calculated with all relevant terms;
    the blue dashed curves are those without the electron-electron Coulomb
    scattering; the green dotted curves are those without the cubic
    Dresselhaus SOC. 
  }
  \label{fig_T}
\end{figure}

First, we compare the calculation via the KSBE approach with the 
experimental data in Ref.~\onlinecite{Muller_08}. In
Fig.~\ref{fig_mag}, we plot the SRT calculated from the KSBEs together with
the experimental data as function of the applied magnetic field. 
Here temperature $T=20$~K, well width $a=16.8$~nm and electron density
$N_e=1.8\times10^{11}$~cm$^{-2}$ as indicated in the
experiment.\cite{Muller_08}
Impurity density $N_i=0.01N_e$ is obtained by fitting the transport
mobility given in Ref.~\onlinecite{Muller_08}. 
It is shown that our calculation agrees well with the experimental data
for various magnetic field strength. 
The fitting gives $\Delta_{\rm R}=0.195$~meV$\cdot${\AA} and
$\gamma_{\rm D}=5.0$~eV$\cdot${\AA}$^3$.
As shown by Sherman,\cite{Sherman_03} the randomness of the
Rashba SOC is related to the effective distance between the QW
conducting sheet and the dopant layer $z_0$. The relationship can be
written as 
\begin{equation}
  \Delta_{\rm R}^2 = \langle\alpha^2\rangle=\zeta^2
  (\frac{e}{\epsilon})^2 \frac{\pi N_e}{2z_0^2},
\end{equation}
where $\zeta=7$~e$\cdot$\AA$^2$ for GaAs.\cite{gate2} 
From the fitting value of $\Delta_{\rm R}$, the corresponding
effective distance $z_0=21$~nm, which is consistent with the sample
parameter given in Ref.~\onlinecite{Muller_08}.\cite{sample}
Our best fitted value of $\gamma_{\rm D}=5.0$~eV$\cdot${\AA}$^3$ is
fairly close to the previous fitting value in GaAs (100) QWs 
($5.7$~eV$\cdot${\AA}$^3$ in Ref.~\onlinecite{Zhou_PRB_07}\cite{factor}  and
$8.6$~eV$\cdot${\AA}$^3$ in Ref.~\onlinecite{wu-exp}) and 
the value reported in {\it ab
initio} calculations ($6.4$, $8.5$~eV$\cdot${\AA}$^3$).\cite{SOC1} 
In fact, the measured values of $\gamma_{\rm D}$ vary from $5.7$ to
  $34.5$~eV$\cdot${\AA}$^3$.\cite{Zhou_PRB_07,list}
This is because the Dresselhaus SOC coefficient  
is affected by the interface inversion asymmetry.\cite{interface}

From Fig.~\ref{fig_mag}, it is also seen that the SRT decreases with
the magnetic field. The underlying physics is as following. 
As first shown by Wu and Kuwata-Gonokami:\cite{Wu_110} in the presence
of an in-plane magnetic field, the in-plane and out-of-plane spins
are mixed and the Dresselhaus spin-orbit field provides an
inhomogeneous broadening which leads to the spin relaxation
and dephasing. Therefore the SRT decreases with $B$. 
For the case with small magnetic field satisfying 
$\omega_L\ll(\gamma_{\|}-\gamma_{\bot})/2$,\cite{criterion} the
effective spin relaxation rate can be written as\cite{Dohrmann_04} 
\begin{equation}
  \gamma_{\rm eff}=\frac{\gamma_{\|}+\gamma_{\bot}}{2}
  - \frac{\gamma_{\|}-\gamma_{\bot}}{2} \sqrt{ 1-\frac{4\omega_L^2}
  {(\gamma_{\|}-\gamma_{\bot})^2}}\;,
  \label{gamma_eff}
\end{equation}
where $\gamma_{\bot}=\Gamma_{\tilde{z}\tilde{z}}$ and
$\gamma_{\|}=\Gamma_{\tilde{y}\tilde{y}}$ with
$\tilde{z}$ and $\tilde{y}$ being the principle axes of the
spin-relaxation-rate tensor $\Gamma$. 
This definition is due to the fact that the $z$-axis and $y$-axis are
not the principle axes of $\Gamma$, when the Rashba and Dresselhaus
SOCs are both present.\cite{Tarasenko,Cartoixa}
It is noted that $\gamma_{\bot}$ differs from $\Gamma_{zz}$ which is
the spin relaxation rate neglecting the Dresselhaus SOC.
The physics of this effect is similar to that of the influence of the
applied magnetic field:\cite{Wu_110} 
the Rashba SOC acts as an in-plane magnetic field, and mixes the
in-plane and out-of-plane spins, thus the Dresselhaus SOC can
affect the out-of-plane spin relaxation.
Our calculation gives the ratio $\gamma_{\|}/\gamma_{\bot}=7.5$,
which is consistent with the huge spin dephasing anisotropy observed
in experiments.\cite{Dohrmann_04,Hagele_05,Muller_08}

In Fig.~\ref{fig_T}, we plot the SRT due to the DP mechanism induced
by the random Rashba field as function of temperature for $B=0$ (a)
and $4$~T (b) with impurity densities $N_i=N_e$ and $0.01N_e$. 
The other sample parameters as well as $\Delta_{\rm R}$ are chosen to
be the same as the previous case.\cite{Ew_tem}
We first concentrate on the case without magnetic field 
[Fig.~\ref{fig_T}(a)]. 
It is interesting to see that the SRT exhibits a peak around the
Fermi temperature $T_{\rm F}=E_{\rm F}/k_{\rm B}$ in the low-impurity-density case.
This peak originates from the electron-electron Coulomb scattering
which dominates the momentum scattering. 
The Coulomb scattering increases with increasing $T$ when $T<T_F$ and decreases
with $T$ when $T>T_F$,\cite{Zhou_PRB_07,Vignale,Ivchenko} 
thus the peak appears around the Fermi temperature.
It is also seen that the SRT increases with increasing $T$ without the Coulomb
scattering. This is because the electron--AC-phonon and
electron--LO-phonon scatterings both increase with temperature monotonically. 
From Fig.~\ref{fig_T}(a), it is also found that the SRT decreases with
increasing $T$ monotonically in the high-impurity-density case. This is because the
electron-impurity scattering dominates the momentum scattering in this case.
Since the impurity scattering depends on temperature 
weakly, the temperature dependence of SRT is mainly determined by
the inhomogeneous broadening from the SOC. With an increase of
temperature, electrons are distributed at higher momentum states. This
leads to the increase of the inhomogeneous broadening and thus a
decrease of the SRT. 

Then we turn to the case with $B=4$~T [Fig.~\ref{fig_T}(b)]. 
For large magnetic field, the SRT is determined
by $\gamma_{\rm eff}=\frac{\gamma_{\|}+\gamma_{\bot}}{2}$. 
In consistence with the anisotropy factor $\gamma_{\|}/\gamma_{\bot}=7.5$,
it is seen that the SRT in this case is three to four times smaller
than that in the case without magnetic field.
The temperature dependence is similar to the case with $B=0$: the SRT
presents a peak around $T_F$ in the low-impurity-density 
case, while decreases with increasing $T$ monotonically in the
high-impurity-density case. The only difference is that the SRT decreases
faster at high temperature in the case with high magnetic field. 
This is because the contribution to the inhomogeneous broadening from
the cubic Dresselhaus SOC becomes significant for large well width and
high temperature.\cite{Weng_04} 
By comparing the SRT without the cubic Dresselhaus SOC, it is found
that the effect of the cubic term increases with $T$ 
faster than the linear Dresselhaus term and the Rashba term (both
are linear). This accelerates the decrease of the SRT at high temperature. 
It is also seen that the SRT without the Coulomb scattering for
$N_i=0$ first increases then decreases and again increases with
increasing $T$. 
This effect is due to the competition of the increase in the
inhomogeneous broadening and the increase in the scattering. 
When $T<50$~K, the contribution from the linear SOC term is
dominant and hence the increase of the inhomogeneous broadening
with rising temperature is slower than that of the electron--AC-phonon and
electron--LO-phonon scatterings. This leads to the increase in the SRT. 
For temperature between $50$~K to $200$~K, 
the contribution of the cubic SOC term is
comparable to that from the linear ones. Thus the effect of the
inhomogeneous broadening increases with increasing $T$ faster than the 
electron--AC-phonon scattering but slower than the
electron--LO-phonon scattering. 
By further noticing that the electron--LO-phonon scattering
surpasses the electron--AC-phonon scattering when $T>100$~K [see the
dashed curve with square in Fig.~\ref{fig_T}(a)], one can understand that
the SRT decreases with increasing $T$ when $50$~K$<T<100$~K but increases 
when $T>100$~K.

In conclusion, we have investigated the SRT due to the DP mechanism
induced by the random Rashba field in symmetric GaAs (110) QWs via the
fully microscopic KBSE approach, where all the relevant scatterings,
especially the electron-electron Coulomb scattering, are explicitly
included.  We show that our calculation is in good agreement with the
experimental data. We also find that the Coulomb scattering makes marked
contribution to the spin relaxation. 
It is predicted that the temperature dependence of the
SRT exhibits a peak in the low-impurity-density case, regardless of
the applied magnetic field. This peak is from the nonmonotonic
temperature dependence of the electron-electron Coulomb scattering.

This work was supported by the National Natural Science Foundation of
China under Grant No.\ 10725417, the National Basic
Research Program of China under Grant No.\ 2006CB922005 and the
Knowledge Innovation Project of Chinese Academy of Sciences.


\begin{thebibliography}{0}
\bibitem{Lorentz} Whether the distribution of $\alpha_{\rm R}$ is
  chosen to be the Gaussian or Lorentzian  distribution function does not
  make any difference to our calculation.
\bibitem{sample} From Ref.~\onlinecite{Muller_08}, it is known
  that the QWs are modulation doped in the middle of barrier layer and
  the width of barrier layer is 80~nm. Since the interaction between
  electron and impurity in the adjacent doping layer is stronger, the
  effective distance can be estimated roughly to be $20$-$40$~nm. 
  Thus the fitting value of $z_0$ we obtain is in the reasonable range.
\bibitem{factor} Due to the different definitions of $\gamma_{\rm D}$, the
 value of $\gamma_{\rm D}$ given here is half of that reported in
  Ref.~\onlinecite{Zhou_PRB_07}.
\bibitem{criterion} This criterion is always satisfied for the cases
  in Fig.~\ref{fig_mag}.
\bibitem{Ew_tem} It is noted that $\Delta_{\rm R}$ also can be
  influenced by temperature. However, since the variation of
  $\Delta_{\rm R}$ with temperature is slower than that of the
  scattering and inhomogeneous broadening in most situation, our 
  calculation with the fixed $\Delta_{\rm R}$ is adequate.
\end{thebibliography}
\end{document}